\documentclass[aps,prl,reprint,preprintnumbers,superscriptaddress,amsmath,amssymb,bibnotes,longbibliography]{revtex4-1}
\usepackage{graphicx}
\usepackage{dcolumn}
\usepackage{bm}

\usepackage{amssymb}
\usepackage{amsmath}
\usepackage{epstopdf}
\usepackage{booktabs}
\usepackage{threeparttable}
\usepackage[pdfborder=001,colorlinks,linkcolor=blue,anchorcolor=blue,citecolor=blue,urlcolor=blue,filecolor=blue,menucolor=blue,runcolor=blue]{hyperref}
\usepackage{lineno}

\begin{document}
\title{CaPtAs: a new noncentrosymmetric superconductor}

\author{W. Xie}
\affiliation{Department of Physics and Center for Correlated Matter, Zhejiang University, Hangzhou 310058, China}
\author{P. R. Zhang}
\affiliation{Department of Physics and Center for Correlated Matter, Zhejiang University, Hangzhou 310058, China}
\author{B. Shen}
\affiliation{Department of Physics and Center for Correlated Matter, Zhejiang University, Hangzhou 310058, China}
\author{W. B. Jiang}
\affiliation{Department of Physics and Center for Correlated Matter, Zhejiang University, Hangzhou 310058, China}
\author{G. M. Pang}
\affiliation{Department of Physics and Center for Correlated Matter, Zhejiang University, Hangzhou 310058, China}
\author{T. Shang}
\affiliation{Physik-Institut, Universit{\"a}t Z{\"u}rich, Winterthurerstrasse 190, CH-8057 Z{\"u}rich, Switzerland}
\author{C. Cao}
\affiliation{Department of Physics, Hangzhou Normal University, Hangzhou 310036, China}
\author{M. Smidman}
\email{msmidman@zju.edu.cn}
\affiliation{Department of Physics and Center for Correlated Matter, Zhejiang University, Hangzhou 310058, China}
\author{H. Q. Yuan}
\email{hqyuan@zju.edu.cn}
\affiliation{Department of Physics and Center for Correlated Matter, Zhejiang University, Hangzhou 310058, China}
\affiliation{Collaborative Innovation Center of Advanced Microstructures, Nanjing University, Nanjing 210093, China}

\date{\today}

\begin{abstract}
We report the discovery of a new noncentrosymmetric superconductor CaPtAs.
It crystallizes in a tetragonal structure (space group $I4_1md$, No.\,109), featuring three dimensional honeycomb networks of Pt-As and a much elongated \emph{c}-axis ($a = b = 4.18 $ \AA{}, and $c = 43.70 $ \AA{}).	
The superconductivity of CaPtAs with $T_c$ = 1.47\,K was characterized by means of electrical resistivity, specific heat, and ac magnetic susceptibility. The electronic specific heat $C_\mathrm{e}(T)/T$ shows evidence for a deviation from the behavior of a conventional BCS superconductor, and can be reasonably fitted by a \emph{p}-wave model. The upper critical field $\mu_0H_{c2}$ of CaPtAs exhibits a relatively large anisotropy, with an in-plane value of around 204 mT and an out-of-plane value of 148 mT. Density functional theory calculations indicate that the Pt-5\emph{d} and As-4\emph{p} orbitals mainly contribute to the density of states near the Fermi level,
showing that the Pt-As honeycomb networks may significantly influence the superconducting properties.

\begin{description}
\item[PACS number(s)]

\end{description}
\end{abstract}

\maketitle

\section{\label{sec:Introduction}Introduction}

 Noncentrosymmetric superconductors (NCSs) have been extensively studied since the discovery of coexistent heavy fermion superconductivity and antiferromagnetism in  CePt$_3$Si \cite{2016Michael,CePt3Si}. Ce\emph{TX}$_3$ (\emph{T} = transition metal, \emph{X} = Si, Ge, Al) are another family of noncentrosymmetric strongly correlated systems \cite{1998CeTPn3,CeTX3,CeZnAl3,CeFeGe3}, which crystallize in the tetragonal BaNiSn$_3$-type structure, and in some cases unconventional superconductivity is observed under pressure in the vicinity of the suppression of antiferromagnetic order  \cite{CeRhSi3,CeIrSi3,CeRhGe3}.  The symmetry-allowed antisymmetric spin-orbit coupling (ASOC) has profound influences on the underlying superconducting state,
  where parity is no longer  a good quantum number due to the breaking of inversion symmetry.
 In principle, the ASOC allows for the admixture of spin singlet and spin triplet pairing states, which has been experimentally observed in Li$_2$Pd$_3$B/Li$_2$Pt$_3$B \cite{2006Yuan,2007GQzheng}, where the lack of electronic correlations allows for the influence of the ASOC to be disentangled from the correlation effects. In addition to the interest in further studying the influence of ASOC on the superconducting pairing, some NCSs are proposed to be candidates for topological superconductivity, such as YPtBi \cite{YPtBi}, BiPd \cite{BiPd}, and PbTaSe$_2$ \cite{PbTaSe2}.

Furthermore, an increasing number of NCSs have been found to show time-reversal symmetry (TRS) breaking upon entering the superconducting state, such as LaNiC$_2$ \cite{LaNiC2}, La$_7$Ir$_3$ \cite{La7Ir3}, Zr$_3$Ir \cite{Zr3Ir}, and a number of Re-based systems with the $\alpha$-Mn structure \cite{Re6Ti, Re6Zr, ReNb, Re24Ti5}. However, in contrast to the previously known TRS breaking strongly correlated  superconductors, such as Sr$_2$RuO$_4$ \cite{1998Sr2RuO4,2000Sr2RuO4,2004Sr2RuO4} and UPt$_3$ \cite{1988UPt3,1993UPt3}, these systems have generally been found to have fully opened superconducting gaps \cite{2016Michael,CJ-LaNiC2,GM-Re6Zr,ReNb,CJ-NbRe,Re24Ti5,Re6Ti,Re6Zr}. The origin of these observations and their relationship with the noncentrosymmetric crystal structures remain to be determined. It is of particular importance to look for new families of NCSs in order to characterize the superconducting properties and examine the influence of a strong ASOC and potential TRS breaking.

 The alkaline-earth based equiatomic ternary compounds have attracted considerable interest since SrPtAs was found to superconduct below 2.4 K \cite{2011SrPtAs}. SrPtAs crystallizes in the KZnAs-type structure (space group $P6_3/mmc$, No.194),  where layers of Pt-As honeycombs and Sr are alternately stacked along the c-axis.~\cite{1986structure1,2011SrPtAs}.
 The global symmetry of SrPtAs is centrosymmetric, however, inversion symmetry is broken within a single Pt-As layer, and it has been proposed that such a locally noncentrosymmetric structure can have significant consequences on the superconductivity, such as allowing singlet-triplet pairing \cite{2012mixed-pairing,2012local_noncentro}. TRS was found to be broken below $T_c$ in SrPtAs \cite{2013TRSB}, and furthermore a chiral \emph{d}-wave pairing has been proposed for the pairing state \cite{2014chiral-d}.

 Recently, two additional examples of alkaline-earth based 111 superconductors were reported, BaPtAs \cite{2018BaPtAs} and BaPtSb \cite{2018BaPtSb}, which also have hexagonal structures with stacked Pt-As/Sb layers. For BaPtAs, three different polymorphs have been found: the SrPtSb-type (space group $P$\={6}$m2$, No.187), YPtAs-type (space group \emph{P}$6_3/mmc$, No.194) and LaIrSi-type (space group $P2_13$, No.198) \cite{2018BaPtAs}.
While the former two types feature ordered honeycomb network structure, and are superconducting below 2.8 K and 2.1-3 K, respectively, the cubic LaIrSi-type BaPtAs was found to be non-superconducting down to 0.1 K \cite{2018BaPtAs}. The ordered honeycomb network structure therefore appears to be favourable for the occurrence of superconductivity.
 In this paper, we report the discovery of superconductivity in CaPtAs with $T_c$ = 1.47~K, which adopts a noncentrosymmetric tetragonal structure consisting of twisted three-dimensional (3D) honeycombs of Pt-As along the $c$-axis~\cite{1986structure2}.

\section{\label{sec:exp} Experimental details}

Polycrystalline samples of CaPtAs were synthesized using a solid state reaction method.
The PtAs$_2$ precursor was synthesized by heating stoichiometric quantities of Pt-powder and As granules at 700 $^\circ$C. Subsequently, Ca, Pt and PtAs$_2$ with a molar ratio of $2:1:1$ were loaded into a crucible and sealed in an evacuated quartz ampule. The ampule was heated up to 1000 $^\circ$C and sintered at this temperature for several days. The sample was then ground and pelleted before being annealed at
950 $^\circ$C for another week. For the single crystal growth, a self-flux method was used. The Ca, Pt and As elements were combined in a molar ratio of 1:1:1, loaded into a crucible and sealed in an evacuated quartz ampule. The ampule was slowly heated up to 1150 $^\circ$C, kept at this temperature for several days before being slowly cooled. The obtained polycrystalline samples are hard and black, while the single crystals are shiny with a purple luster and irregular plate shape [inset of Fig.~\ref{fig1}(c)]. The crystal structure of the synthesized samples werechecked by powder x-ray diffraction (XRD) measured on a PANalytical X'Pert MRD powder diffractometer with Cu $K_{\alpha1}$ radiation monochromated by graphite. For single crystals, the orientation was also confirmed by XRD. The chemical composition was determined by energy-dispersive x-ray spectroscopy (EDS) using a Hitachi SU-8010 Field Emission scanning electron microscope. The electrical resistivity and heat capacity measurements were carried out using a Quantum Design Physical Property Measurement System (PPMS-9 T). A four-contact method was used for resistivity measurements. Alternating-current (ac) magnetic susceptibility was measured in a $^3$He refrigerator down to 0.3\,K using an $ac$ magnetometer.

Electrical resistivity measurements under hydrostatic pressures were carried
out using a piston-cylinder clamp-type cell. Daphne oil 7373 was used as a pressure transmitting
medium. The pressure was determined by measuring the superconducting transition of pure Pb.

DFT calculations were performed using the Vienna \emph{ab-initio} simulation package (VASP). In particular, the projected augmented wave (PAW) method and Perdew-Burke-Ernzerhoff (PBE) flavor of generalized gradient approximation were employed. A plane-wave basis up to 500 eV and 12$\times$12$\times$12 $\Gamma$-centered K-mesh were used to ensure convergence. Both the lattice constants and the atomic internal coordinates were optimized so that the forces on each atom were smaller than 0.01 eV/{\AA} and the internal stress smaller than 0.1 kbar.

\section{\label{sec:results} Results and discussion}

\subsection{Structure and composition}
\begin{figure}[tb]
     \centering
     \includegraphics[width=1.0\columnwidth,keepaspectratio]{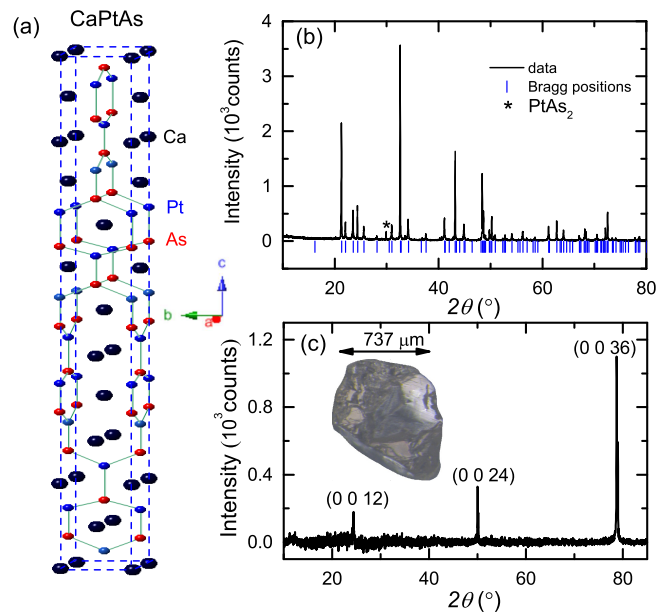}
     \caption{(Color online) (a) Crystal structure of CaPtAs where one unit cell is displayed. The black, blue and red balls correspond to Ca, Pt, and As atoms, respectively. (b) Room temperature powder x-ray diffraction (XRD) pattern of polycrystalline CaPtAs. The vertical bars indicate the Bragg peak positions. The unindexed peaks marked by an asterisk are from nonmagnetic and nonsuperconducting PtAs$_2$. (c) XRD pattern collected by measuring a piece of single crystal, a photo of which is shown in the inset. }
     \label{fig1}
\end{figure}

The crystal structure of CaPtAs is shown in Fig.~\ref{fig1}(a) \cite{1986structure2}. Both Ca, Pt, and As atoms occupy the 4\emph{a} sites in the unit cell with $a = b = 4.18$ \AA{}, and $c = 43.70$ \AA{}. Different from the other alkaline-earth 111-type compounds, where the honeycomb networks layers are stacked along the \emph{c}-axis, the Pt-As formed honeycomb-like networks in CaPtAs are aligned (with the normal direction of the hexagons) along the \emph{a} (or \emph{b}) axis, where the orientation of the hexagons is rotated by 90$^\circ$ upon translating along the c-axis.
As a result, the \emph{c}-axis is extensively elongated, leading to a long unit cell where the lattice parameter \emph{c} is over 10 times longer than the in-plane value. 
Also due to such twisted structural feature, there is a lack of an inversion center along the \emph{c}-axis.

Powder XRD data for CaPtAs are shown in Fig.~\ref{fig1}(b). Almost all the diffraction peaks can be indexed by a tetragonal structure with space group of $I4_1md$ (No.109). Peaks corresponding to a small amount of an additional phase PtAs$_2$, which is not superconducting, were also detected. The XRD pattern measured on a CaPtAs single crystal is shown in Fig.~\ref{fig1}(c). Only the (00$l$) reflections are present, indicating a well-defined crystalline orientation and that the $c$-axis is perpendicular to the measured crystal plane.
EDS shows the chemical composition of the crystal is Ca:Pt:As = 32.79: 33.76: 33.95, being close to the stoichiometric ratio of 1: 1: 1.

\subsection{Zero-field superconducting properties}

\begin{figure}[tb]
\centering
\includegraphics[width=1.0\columnwidth,keepaspectratio]{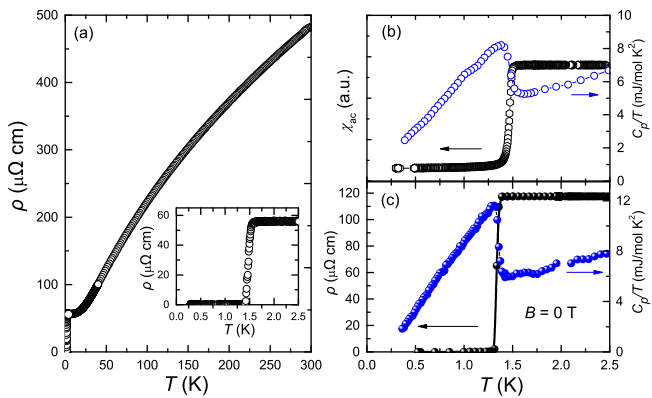}
\caption{(Color online) Temperature dependence of the electrical resistivity $\rho(T)$ of polycrystalline CaPtAs in zero field from 0.3 K to 300 K. The inset shows the low temperature behavior in the vicinity of $T_c$. (b) The ac susceptibility $\chi_{ac}(T)$ (in phase part) and specific heat as $C_\mathrm{p}/T$ for polycrystalline CaPtAs at low temperature. (c) Low temperature $\rho(T)$ and $C_\mathrm{p}/T$ for single crystal CaPtAs. }
\label{fig2}
\end{figure}

Figure~\ref{fig2} shows the zero-field characterization of superconductivity in CaPtAs, including the electrical resistivity $\rho(T)$, ac magnetic susceptibility $\chi_\mathrm{ac}(T)$, and specific heat $C_\mathrm{p}/T$. $\rho(T)$ of polycrystalline CaPtAs between 0.3 and 300\,K is shown in Fig.~\ref{fig2}(a). The resistivity decreases upon decreasing the temperature, showing typical metallic behavior. No phase transition related to the magnetic, structural, or charge density wave can be found before the superconducting transition at $T_c$ = 1.47\,K, which is defined from where $\rho(T)$ drops to 50$\%$ of the normal state value [inset of Fig.~\ref{fig2}(a)]. The superconductivity of polycrystalline CaPtAs was further confirmed by measuring specific heat and ac magnetic susceptibility [see Fig.~\ref{fig2}(b)]. In $C_\mathrm{p}/T$, $T_c$ was determined by the balanced-entropy method, and in $\chi_\mathrm{ac}(T)$, it was defined as the midpoint of the transition. The values are 1.47 K and 1.475 K, respectively.
For measurements of single crystals, as shown in Fig.~\ref{fig2}(c), the $T_c$ values (1.34 K from $\rho(T)$ and 1.35 K from $C_\mathrm{p}/T$) are slightly smaller than those for the polycrystalline sample, but the superconducting transition is much sharper. The transition widths in the resistivity $\Delta$$T$, defined as the difference between the onset of the transition and where $\rho = 0$ (= $T_c^{onset}-T_c^{zero}$), are found to be 0.16\,K and 0.08\,K, for polycrystalline and single crystal samples, respectively.

\begin{figure}[tb]
     \begin{center}
     \includegraphics[width=1.0\columnwidth]{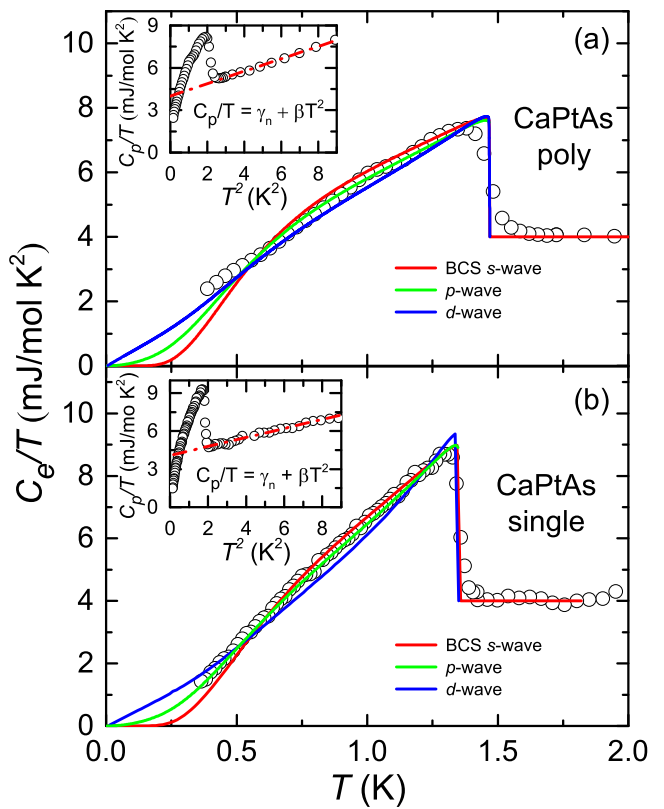}
     \end{center}
     \caption{(Color online) Electronic contribution to the heat capacity as $C_e(T)/T$ for (a) polycrystalline and (b) single-crystalline CaPtAs. The solid red, green, and blue lines in both (a) and (b) are results from fitting using a single band \emph{s}-, \emph{p}-, and \emph{d}- wave model, respectively. The fitted values of $\Delta_0$ for \emph{s}-,  \emph{p}-, and \emph{d}- wave model is 1.4, 1.7, and 2 $k_BT_c$, respectively in (a), and 1.6, 2, and 2.4 $k_BT_c$ in (b). Insets show $C_\mathrm{p}(T)/T$ versus $T^2$ and the corresponding fitting with $C_\mathrm{p}/T=\gamma_\mathrm{n}+\beta T^2$.}
     \label{fig3}
\end{figure}

Figure~\ref{fig3} displays the electronic contribution to the specific heat ($C_\mathrm{e}/T$) of both polycrystalline and single-crystalline CaPtAs. As shown by the dashed-lines in the insets,
the phonon contributions are estimated by fitting the normal-state specific heat 
to $C_\mathrm{p}/T = \gamma_\mathrm{n} + \beta T^2$, where $\gamma_\mathrm{n}$ is the normal-state electronic specific heat coefficient, and $\beta T^2$ corresponds to the low temperature phonon contribution.
The estimated $\gamma_\mathrm{n}$ is 4\,mJ/mol-K$^2$ for both polycrystal and single crystal samples, while their $\beta$ values are 0.44\,mJ/(mol-K$^4$) and 0.33\,mJ/(mol-K$^4$), respectively.

The Debye temperatures $\Theta_\mathrm{D}$ can be calculated according to $\Theta_\mathrm{D}=\sqrt[3]{\frac{12\pi^4\mathrm{nR}}{5\beta}}$, where \emph{n} = 3 is the number of atoms per formula unit and \emph{R} is the molar gas constant \cite{Kittel}. The $\Theta_\mathrm{D}$ are 237\,K and 260\,K for polycrystals and single crystals, respectively. Based on the McMillan formula~\cite{McMillan}, we estimated the respective electron-phonon couplings to be $\lambda_{ep} \approx 0.5$ and 0.48, using $\lambda_{ep} = (1.04 + \mu^* \log[\Theta_\mathrm{D}/(1.45*T_c)])/((1-0.62*\mu^*)\log[\Theta_\mathrm{D}/(1.45*T_c)]-1.04)$, where the repulsive screened coulomb parameter $\mu^*$ is taken to have the typical value of 0.13. The relatively small $\lambda_{ep}$ indicates that CaPtAs is a weakly coupled superconductor. In addition, $\gamma\rm_n$ is related to the density of states at the Fermi level $N(E_F)$ by $\gamma_\mathrm{n}=\frac{\pi^2}{3}k^2_BN(E_F)$,  where $k_B$ is the Boltzmann constant. $N(E_F)$ was calculated to be 1.7 states/(eV atom).
The electronic specific heat jump 
$\Delta C/\gamma_\mathrm{n} T_c$ was determined from Fig.~\ref{fig3}, which is about 0.92 for the polycrystalline sample, and 1.25 for the single crystal. Both values are smaller than the BCS value. Such reduced $\Delta C/\gamma_\mathrm{n} T_c$ could be caused by effects such as multiple gaps or gap anisotropy~\cite{2001MgB2, 1996Nb3X4}, or may be due to extrinsic effects such as non-superconducting regions of the sample.

The temperature-dependent electronic heat-capacity 
in the superconducting state can be calculated from
$C_\mathrm{e} = T\frac{dS}{dT}$, where the superconducting contribution to the entropy $S(T)$ was calculated using~\cite{GL-CL}

\begin{equation}
S(T)= -\frac{6\gamma\rm_n}{\pi^2k_B}\int_0^\infty[f\ln f+(1-f)\ln(1-f)]d\epsilon,
\label{equation1}
\end{equation}

\noindent in which $f=(1+e^{E/k_BT})^{-1}$ is the Fermi function, $E(\epsilon)=\sqrt{\epsilon^2+\Delta^2(T)}$ is the excitation energy of quasiparticles, where $\epsilon$ are
the electron energies measured relative to the chemical potential \cite{GL-CL,1973QP-energy}, and $\Delta(T)$ is the temperature dependent gap function, which in the BCS \emph{s}-wave model can be approximated as
\begin{equation}
\Delta_s(T) = \Delta_0 \tanh[1.82[1.018(T_c/T-1)]^{0.51}],
\label{equation2}
\end{equation}
where $\Delta_0$ is the superconducting gap at zero temperature. While in the \emph{p}-, and \emph{d}- wave model, $\Delta(T)$ is $\Delta_s(T) \sin{\theta}$ and $\Delta_s(T) \cos{2\phi}$, respectively, with the existence of point node and line node in the respective gap function.

The results from fitting with the above-mentioned models are shown by the solid lines in Fig.~\ref{fig3}, and the fitted $\Delta_0$ for each model are listed in the caption of Fig. \ref{fig3}. It seems that $C_\mathrm{e}(T)/T$ of polycrystalline sample can not be well fitted by any of the three models. While for the single crystal data in Fig. \ref{fig3}(b), we find that the \emph{p}-wave model can fit $C_\mathrm{e}(T)/T$ across the measured temperature, while the \emph{s}-wave and \emph{d}-wave models show different degrees of deviation from the data. However, in order to clarify the structure of the superconducting gap and the nature of the pairing state, measurements using high quality sample to lower temperatures are required, in particular to determine whether there are nodes in the gap.

\subsection{In-field superconducting properties}
\begin{figure}[tb]
     \begin{center}
     \includegraphics[width=1.0\columnwidth,keepaspectratio]{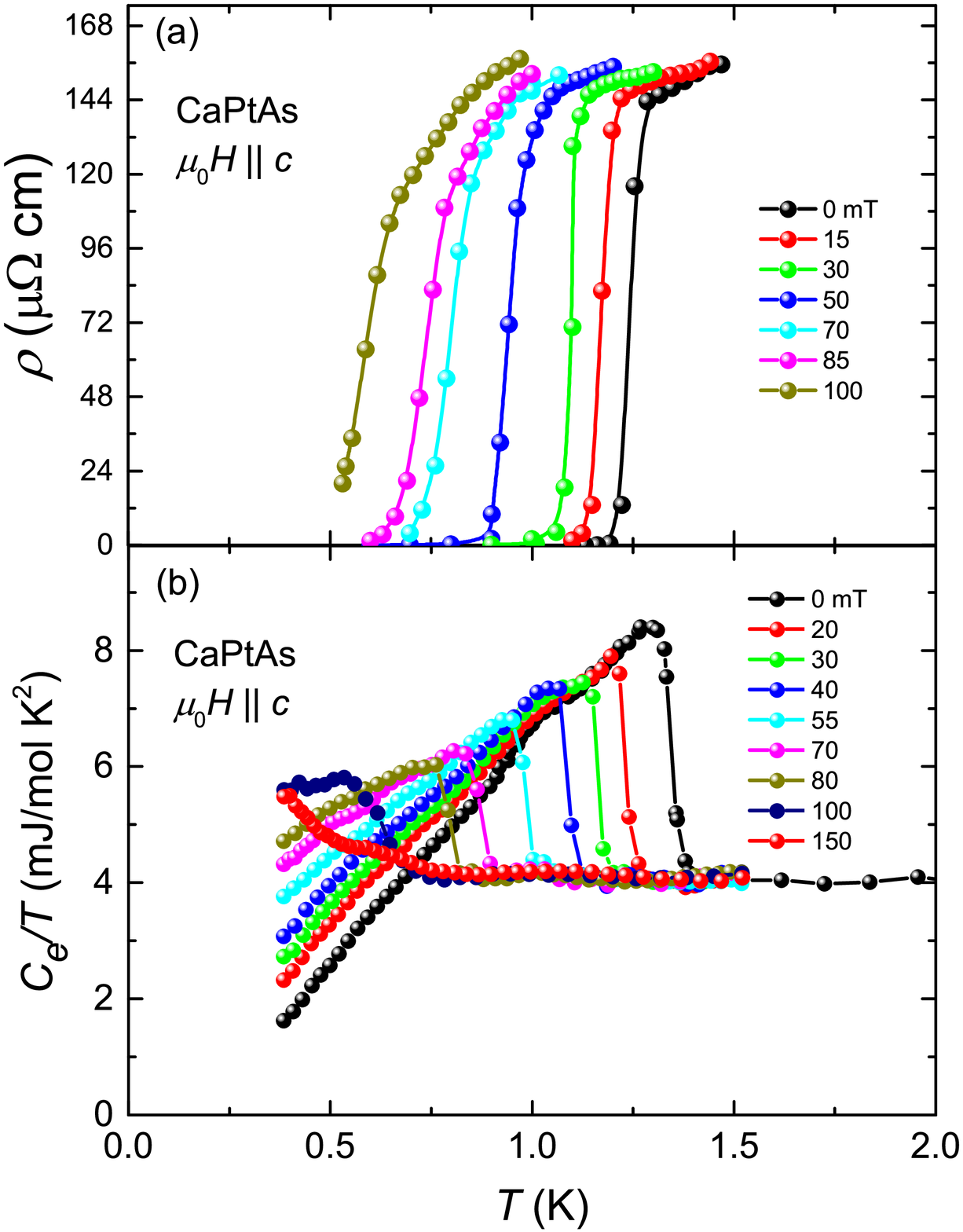}
     \end{center}
     \caption{(Color online) The temperature dependence of (a) $\rho(T)$ and (b) $C_\mathrm{e}(T)/T$ for single crystalline CaPtAs under various magnetic fields, where $\mu_0H \parallel c$.  }
     \label{fig4}
\end{figure}

\begin{figure}[tb]
     \begin{center}
     \includegraphics[width=1.0\columnwidth,keepaspectratio]{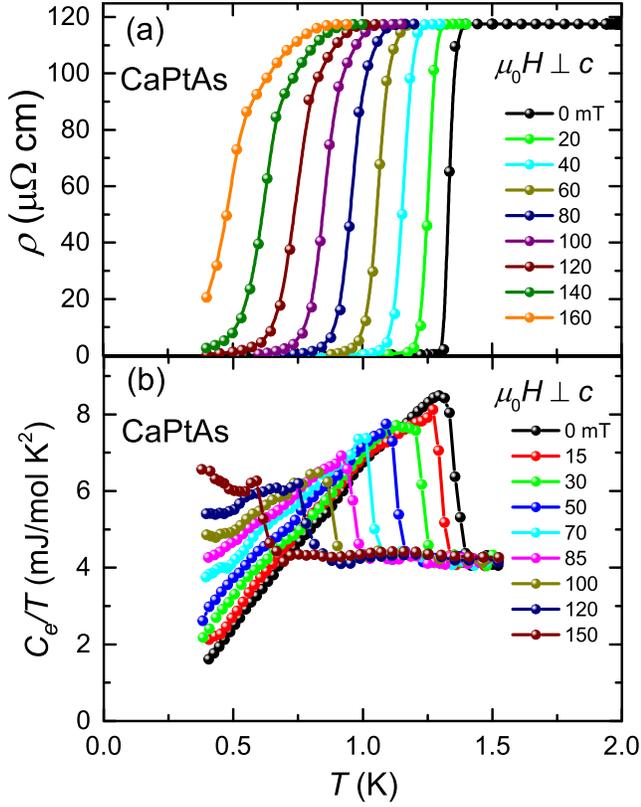}
     \end{center}
     \caption{(Color online) The temperature dependence of (a) $\rho(T)$ and (b) $C_e(T)/T$ for single-crystalline CaPtAs under various magnetic fields, with $\mu_0H \perp c$.}
     \label{fig5}
\end{figure}

To study the upper critical field $\mu_0H_{c2}$ of CaPtAs, we performed electrical resistivity and heat capacity measurements under various magnetic fields up to 0.2\,T. 
Considering the largely elongated \emph{c}-axis in the lattice, we focus on the anisotropic of $\mu_0H_{c2}$, which was checked by applying magnetic fields along the \emph{c}-axis and within the \emph{ab}-plane. The data are shown in Fig.~\ref{fig4} and Fig.~\ref{fig5}. Upon applying the magnetic field, the transition is suppressed to lower temperatures and becomes broader. In both Fig.~\ref{fig4}(b) and Fig.~\ref{fig5}(b), $C_\mathrm{e}(T)/T$ displays a low temperature upturn at higher magnetic fields, which we ascribe to a nuclear Schottky contribution.


\begin{figure}[tb]
     \begin{center}
     \includegraphics[width=1.0\columnwidth,keepaspectratio]{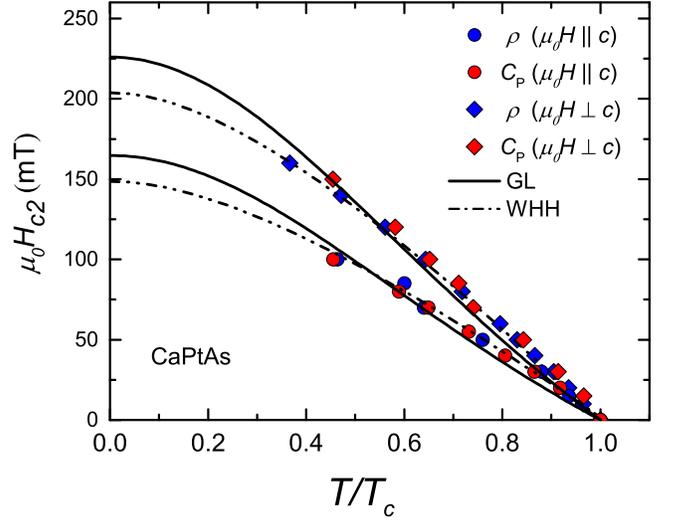}
     \end{center}
     \caption{(Color online) Upper critical field $\mu_0H_{c2}$ versus the normalized temperature $T/T_c$. The magnetic fields were applied along both the \emph{c}-axis and \emph{ab}-plane, respectively. The fitted curves using the WHH (dashed line) and GL (solid line) model are also displayed.}
     \label{fig6}
\end{figure}

We summarize the results in Fig. \ref{fig6}, by plotting the upper critical fields $\mu_0H_{c2}$ versus 
the normalized critical temperature $T/T_c$, where $T_c$ is the critical temperature under zero-field. For the resistivity, $T_c$ was determined from the temperature where $\rho(T, H)$ dropped to half of the normal state value, while $T_c$ from $C_\mathrm{e}(T, H)/T$ was determined from the position of the midpoint of the jump at the transition.
 It can be seen that for both field orientations, the data points from resistivity and specific heat coincide well. This is rather different with that observed for PdTe$_2$ \cite{2017PdTe2,2019PdTe2} and PbTaSe$_2$ \cite{2016PbTaSe2,2019PbTaSe2}, where the differences were attributed to possible surface contributions. The fitted $\mu_0H_{c2}$(0) of CaPtAs is smaller than that of BaPtAs, where the extrapolated $\mu_0H_{c2}$(0) is 0.55 T for the SrPtSb-type structure, and 3 T (or 0.5 T) for the YPtAs-type phase \cite{2018BaPtAs}.

 Both the Ginzburg-Landau (GL) model \cite{GL-theory} and Werthamer-Helfand-Hohenberg (WHH) model \cite{WHH} were applied to analyze the upper critical field. The obtained zero temperature values are summarized in Table \ref{table1}. For the WHH model analysis, the Maki parameter $\alpha_M$ is found to be near zero. The Pauli limit \cite{Pauli-limit} of $\mu_0H_{c2}$ is estimated to be 1.86$T_c$ = 2.7 T (using $T_c$ of 1.47 K), which greatly surpasses the observed values. Therefore, it is clear that CaPtAs is an orbital limited superconductor. By using the value of $\mu_0H_{c2}(0)$ from the WHH model shown in Table \ref{table1}, we calculated the Ginzburg-Landau coherence length \cite{GL-CL} to be 472\,\AA{} for $\mu_0H \parallel c$ and 402\,\AA{} for $\mu_0H \perp c$, using $\xi_\mathrm{GL}=(\phi_0/2\pi\mu_0H_{c2})^{1/2}$, where the $\phi_0$ is the flux quantum. Also by using $l = 1.27\times10^4\times(\rho_0 n^{2/3}S/S_F)^{-1}$ \cite{1979mean-free-path}, we estimated the averaged mean free path as $l = 15\,\AA{}$, where we use a residual resistivity $\rho_0$ of 120 $\mu\Omega$ cm [Fig. \ref{fig2}(c)], a  carrier concentration \emph{n} of about 1.8$\times10^{22}$ cm$^{-3}$ (from the measured Hall coefficient at 2 K), and taking the ratio of Fermi surface area $S/S _F$ $\approx$ 1. The BCS coherence length $\xi_0$ was estimated to be 3910$\,\AA{}$, using $\xi_0=7.95\times10^{-17}\times(n^{2/3}S/S_F)(\gamma_\mathrm{n} T_c)^{-1}$~\cite{1979mean-free-path}. Comparing $\xi_0$ with \emph{l}, one can conclude that the single crystal is in the dirty limit.

It can be seen that the experimental data can be better fitted by the WHH model, as shown in Fig.~\ref{fig6}. For both models, the $\mu_0H_{c2}(0)$ in the ab-plane is moderately larger than that along the \emph{c}-axis. Since in the dirty limit, $\xi_\mathrm{GL} = 0.855(\xi_0l)^{1/2}$, where $\xi_0 = \hbar\nu_F/\pi\Delta(0)$~\cite{GL-CL}, it is possible that the anisotropy of $\mu_0H_{c2}(0)$ comes from anisotropy of the Fermi velocity $\nu_F$, gap anisotropy or mean free path.

\begin{table}[tb]
\footnotesize
\caption{The fitting results of $\mu_0H_{c2}(0)$ using the GL and WHH models.}\label{tab1}
\doublerulesep 0.1pt \tabcolsep 11pt 
\begin{tabular}{cccc}
\toprule
   & GL-model  &  WHH-model \\

    \hline

     \raisebox{0ex}{$\mu_0H\parallel c$} & $\mu_0H_{c2}(0)$ = 164(1)mT & $\mu_0H_{c2}(0)$ = 148(1)mT \\

    \hline

    \raisebox{0ex}{$\mu_0H \perp c$ } & $\mu_0H_{c2}(0)$ = 226(1)mT & $\mu_0H_{c2}(0)$ = 204(1)mT  \\
\bottomrule
\end{tabular}
\label{table1}
\end{table}

\subsection{Pressure effect and DFT calculations}

\begin{figure}[tb]
     \begin{center}
     \includegraphics[width=1.0\columnwidth,keepaspectratio]{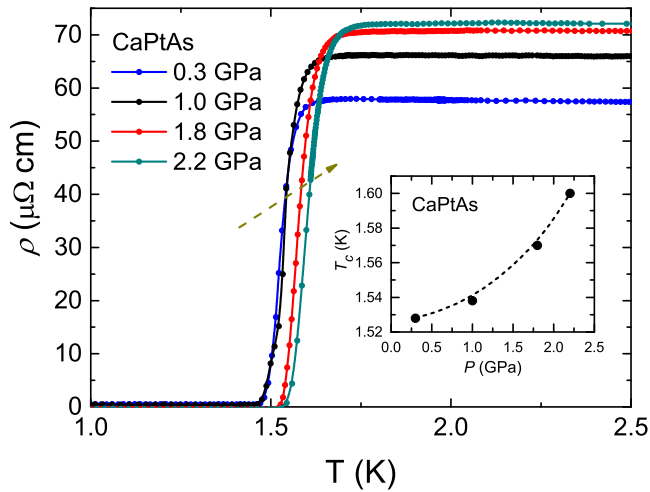}
     \end{center}
     \caption{(Color online) Low temperature $\rho(T)$ under pressures up to 2.2 GPa, showing a slight increase of $T_c$. The inset shows the pressure dependence of $T_c$. }
     \label{fig7}
\end{figure}

The constituent elements of CaPtAs are close to the alkaline-earth based 111 compounds, such as SrPtAs, BaPtAs. However, its structure is more analogous to LaIrAs \cite{2014LaMP}, which also corresponds to the noncentrosymmetric tetragonal space group $I4_1md$ (No.109). In CaPtAs and LaIrAs, the Pt/Ir-As atoms form three dimensional honeycombs networks. In particular, the honeycombs in CaPtAs are rotated upon translating along the \emph{c}-axis, giving rise to the long periodicity. Such a 3D honeycomb structure is also found in MgPtSi \cite{2015MgPtSi}, where the Pt-Si atoms form buckled honeycomb layers with chemical bonds between layers.
 In addition, MgPtSi was proposed to be near a structural instability, based on the observation that pressurized CaPtSi (3-4 GPa) has the same orthorhombic structure as that of MgPtSi while CaPtSi is cubic at ambient pressure \cite{2015MgPtSi}. Considering the different structures of BaPtAs, SrPtAs and CaPtAs, we also checked the electronic properties of CaPtAs by applying hydrostatic pressure. As shown in Fig. \ref{fig7}, there is a small increase of $T\rm_c$ and residual resistivity $\rho_0$ under applied pressures up to 2.2 GPa, while no additional phase transitions are observed. According to the BCS formula, the slight increase of $T_c$ under pressure is likely attributed to the increase of Debye temperature, but a change of $N(E_F)$ may also affect $T_c$. Whether this is applicable to CaPtAs remains to be determined.

Among the 111 superconductors discussed above, only SrPtAs has been extensively characterized, which shows evidences for unconventional superconductivity from the presence of time reversal symmetry breaking \cite{2013TRSB}. While for the others, evidence for conventional BCS-type superconductivity has been reported. For example, the $T_c$ of BaPtSb can be successfully estimated from the first-principle theory considering the electron-phonon interaction \cite{2019BaPtSb}, and $T_c$ of both MgPtSi and LaIrAs can be well accounted for by considering $N(E_F)$ and the electron-phonon coupling strength \cite{2015MgPtSi, 2014LaMP}.

\begin{figure}[tb]
     \begin{center}
     \includegraphics[width=1.0\columnwidth,keepaspectratio]{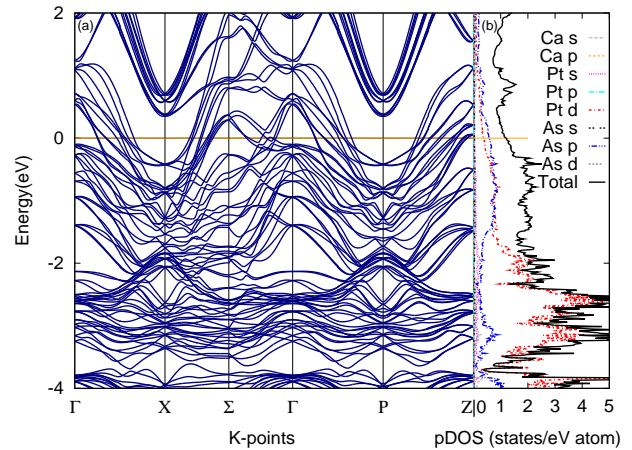}
     \end{center}
     \caption{(Color online)(a) Band structure and (b) partial density of states (pDOS) of CaPtAs obtained from DFT calculations taking into account SOC.}
     \label{fig8}
\end{figure}

 We have also performed DFT calculations for CaPtAs. Our results show that the band structure is complex due to the large number of atoms per unit cell, as shown in Fig. \ref{fig8}. The total $N(E_F)$ amounts to 1.18 states/(eV atom), which is slightly less than that derived from the experimental $\gamma$ value. This could be due to the presence of electron-phonon and/or electron-electron interactions in CaPtAs. It can be seen that the contributions to $N(E_F)$ are mainly from Pt-5\emph{d} and As-4\emph{p} orbitals. This indicates that the 3D Pt-As honeycomb structure is important for determining the superconducting properties in CaPtAs, which is analogous to that observed in MgPtSi \cite{2015MgPtSi} and LaIrAs \cite{2014LaMP}. The inclusion of spin-orbit coupling (SOC) in the calculations gives rise to an ASOC, which leads to a momentum dependent band splitting of around 50-100 meV. This corresponds to a moderately large ASOC effect compared to other reported NCSs \cite{2016Michael}.
 For MgPtSi and LaIrAs, although there is a large SOC arising due to Pt/Ir, the parity-mixing effect and spin triplet contribution seems to be negligible \cite{2015MgPtSi, 2014LaMP}.
 In CaPtAs, the electronic specific heat $C_\mathrm{e}/T$ shows a deviation from a single gap \emph{s}-wave model, but can be well fitted by a \emph{p}-wave model, which raises the possibility of unconventional pairing. However, additional measurements at lower temperatures are needed to clarify the gap structure and whether there is nodal superconductivity. Note that TRS breaking for CaPtAs has also been detected by our preliminary muon spin relaxation measurements, which will be published elsewhere.

\section{Summary and conclusion}\label{sec:4}
To summarize, we have successfully synthesized both poly- and single- crystalline samples of a new superconductor CaPtAs. Unlike SrPtAs and BaPtAs, CaPtAs crystallizes in a body-centered tetragonal noncentrosymmetric structure consisting of Pt-As honeycombs which extending along a much elongated \emph{c}-axis.

Zero-field resistivity, ac magnetic susceptibility and specific heat measurements demonstrate the presence of
bulk superconductivity in CaPtAs with a $T_c$ of 1.47~K. The specific heat jump at $T_c$, i.e., $\Delta$\emph{C}/$\gamma$$T_c$ is around 0.92--1.25, which
is smaller than the value of 1.43 from weak-coupling BCS theory. The temperature dependence of the electronic specific heat can not be well fitted by a single gap BCS model.
The upper critical field shows a moderate anisotropy, with zero temperature values of 204 mT for $H \perp c$, and 148 mT for $H \parallel c$. Applying hydrostatic pressure up to 2.2 GPa leads to a slight enhancement of the $T_c$, and residual resistivity $\rho_0$. From the DFT calculations, we found that the ASOC-induced band splitting is around 50-100 meV. Furthermore, the DOS near Fermi level is dominated by Pt-5\emph{d} and As-4\emph{p} orbitals, suggesting that an important role is played by the Pt-As honeycombs in influencing the superconducting properties.

Although the specific heat could be reasonably fitted by a \emph{p}-wave model, additional measurements at lower temperatures sensitive to low energy excitations, such as the magnetic penetration depth, are necessary to characterize the gap structure and determine the nature of the pairing state.

\section{Acknowledgments}
This work was supported
by the National Key R$\&$D Program of China (Grants No. 2016YFA0300202 and No. 2017YFA0303100),
the National Natural Science Foundation of China (Grants No. U1632275 and No. 11874320), and the Fundamental Research Funds for the Central Universities.

\end{document}